\documentclass[a4paper,11pt]{article}
\pdfoutput=1 

\usepackage{jinstpub} 

\title{\boldmath Digitized Waveform Signal Processing for Fast Timing: An Application to SiPM Timing in the Presence of Dark Count Noise}

\author[a,1]{S.~White\note{Corresponding author.}\note{Also at University of Virginia}}
\author[b]{A.~Heering}

\emailAdd{sebastian.white@cern.ch}

\affiliation[a]{European Organization for Nuclear Research (CERN)\\ CH-1211 Geneve 23, Switzerland}
\affiliation[b]{University of Notre Dame\\ Notre Dame, Indiana, USA}

\abstract{
In this paper we illustrate techniques for digitized waveform signal processing of fast timing detectors.
In the example discussed here, timing analysis of SiPM signals in the presence of high Dark Count Rates, 
a large data set of digitized waveforms is used to develop an optimal strategy relevant to the electronics 
front end design.

}


\proceeding{Innovative Particle and Radiation Detectors 2019 - IPRD19\\
  14$^\text{th}$ - 17$^\text{th}$ October 2019\\
  Siena}

\begin{document}
\maketitle
\flushbottom
\section{Introduction}
\label{sec:Introduction}

	In an earlier note\cite{Uli} we discussed techniques for signal processing from detectors whose data is presented in the form of discrete time samples. 
	
	In collider experiments operating with bunched beams such data naturally occur when data are pipelined in digital form to enable a delayed L1 trigger. In the case of calorimeter data (eg ATLAS LAr EM Barrel) 5 or so samples may be available in the output stream permitting a subsequent optimization of the signal processing (eg. favoring energy rather than time resolution \cite{Cleland}).
	
	As the Large Hadron Collider luminosity continues to climb beyond the original design value, the events captured in a single frame (ie bunch crossing) contain a random overlap of physics objects from interactions captured in the same frame with increasing probability. This pileup induced background can be mitigated by proper association of these objects with their reconstructed interaction points- in both space and time-\cite{intro}\cite{seb}. Since the time frame of an LHC bunch crossing is of order 170 picoseconds rms, the Phase II upgrades are targeting particle time resolutions of order 10s of picoseconds(ps).
	
	R$\&$D on detectors and electronics to achieve this goal has benefitted from the availability of oscilloscopes and dedicated waveform digitizers that can capture data at $\sim 20$ GSa/s or greater permitting the evaluation of different signal processing algorithms tailored to the actual device and conditions under test.
	Rather than committing to a given timing circuit for detector development as was usually the case in the past, this approach can be useful in developing different algorithms before committing to a particular ASIC design, for example.
	In the following we illustrate this approach for the case of Geiger Mode avalanche photodetectors (aka SiPM) which will be affected during their use at High Luminosity LHC by the high integrated radiation field (eg.
$\sim 10^{13}$ neq/cm$^2$ in the below example) with a consequent increase in dark count rate (DCR) of several orders of magnitude.
	
\section{ Optimal Filter}
\label{sec:Optimal}
	The particular use case we have in mind is the CMS MIP Timing Detector barrel, wherein SiPMs will be coupled to thin (~3 mm) LYSO bars to detect the light signal produced by a relativistic charged particle traversing the bar. 
	The readout electronics must produce a time of arrival (TOA) signal which compensates for the variable pulse amplitude. This so-called walk compensation is already familiar in SiPM applications (eg in medical imaging) where DCR is not an issue.
	We may then ask how best to maintain SiPM time resolution in the presence of DCR. Does optimal processing change as a function of DCR noise rate? For example, is there a single optimal bandpass selection over a large range of DCR rate? 
	
	The background count rate effectively introduces an instability of the baseline (in proportion to the SiPM recovery time) ahead of the leading edge of the signal to be timed. 
	While there are a number of techniques to accomplish baseline restoration, the most common is to sample the baseline just ahead of the signal and subtract this value from subsequent samples. Normally this is equivalent to a high pass filter since the baseline jitter is dominated by lower frequencies than those characteristic of our signal ( $\sim 300$ MHz in the example discussed below).
	
	The actual baseline restoration algorithm introduced below is accomplished by summing the SiPM signal with a delayed ($\delta$t) and inverted copy as in eqn 4.1. The conclusion of the study presented below is that this DCR noise filter algorithm effectively reduces the impact of DCR on the time resolution (by roughly a factor of 2 in our example) with little sensitivity to the value of $\delta$t used in the filter. On the other hand the optimal timing threshold does vary and should be configurable over the expected range of DCR level.
\section{Method}
\label{sec:Method}

	Measurements were performed in a dark room (LHCb fiber lab) with an LED head -model PLS-8-2-635(497 nm)- driven by a PicoQuant Model PDL800 laser driver. 
	An LED (DC operation peak emission 470 nm) attached to a variable current source placed near the SiPM under test 
was used to vary artificially the dark count rate in the case of un-irradiated SiPMs. 

We found that the minimum time jitter (so small that no correction is applied for it) relative to the trigger was obtained when using the PicoQuant internal trigger, which for the measurements reported here had a trigger rate of 1 MHz/32=31.25 kHz.
For the bulk of these measurements the SiPM was operated at 2 V overvoltage and had an internal gain of 1.8$\times 10^5$. Based on the known gain we could cross-reference the expected mean light intensity (ie the mean number of photoelectrons, $< N_{pe} >$)
per pulse to the SiPM bias current. Alternatively the same could be done for the dark count rate as we increased the ambient light and hence the bias current (since the cross-talk was of order few percent so 1 dark count=1 photoelectron).

	The spec for the PicoQuant LED head lists a light pulse time spread 350 ps (rms) but we have not yet measured directly the photon time distribution within the pulse.

	While this time distribution is obviously not the same as the LYSO output signal produced by charged particles in the MIP timing detector, it is nevertheless a useful
model for determining the functional form of the DCR resolution term. This model also could be easily used to confirm the simulations currently used for estimating the LYSO layer timing.
	
	The actual setup is shown in Figure 1. In this photo an HPK S12572 - 015 SiPM is mounted on a discrete transimpedance amplifier (designed to have similar impedance and bandwidth as the planned ASIC to be used in CMS)
board. 
For most of the data discussed in this report the amplifier was housed in an aluminum box where its 300 mW heat was dissipated by placing the aluminum box in good thermal contact with a fan cooled large aluminum plate. The ambient temperature was measured at all times to be between 20-21 $^{\circ}$C.
When operating the SiPM with 40 GHz DCR it will dissipate about 70 mW of self heating at 2 V overvoltage. To keep the temperature of the SiPM to within 1-2 $^{\circ}$C of 20 degrees the SiPM itself was housed in a 5x5 mm ceramic package that was put in good thermal contact with the aluminum box by using a 0.5 mm adhesive gap filler with a thermal conductivity of 1 W/mK.

\begin{figure}
\centering
\centerline{\includegraphics[width=7cm, height=7cm]{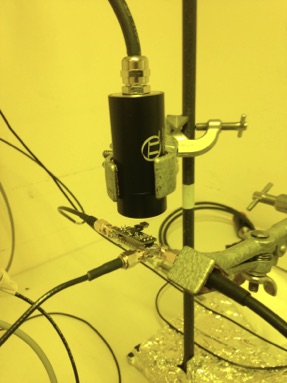}}
\caption{Early version of the setup with bare TIA amplifier and SiPM positioned below LED head. For the bulk of measurements the SiPM was temperature stabilized with a heat sink/fan. }
\label{fig:shannon}       
\end{figure}

\section{Signal Processing}
\label{sec:SigProc}

  	Both trigger (output from PicoQuant) and SiPM  waveforms were recorded on a Lecroy-Teledyne digital scope with 1 GHz analog bandwidth and a sampling frequency of 20 GSa/s.
Since the area around building 4 at CERN has a high level of (GSM?) noise, it was useful to develop a digital bandpass filter to pre-process the waveforms \cite{Uli}. A 500 MHz low pass filter was effective in eliminating this RF noise without
any degradation of the signal. The effect of this filter is shown in Figure 2.

\begin{figure}
\centering
\centerline{\includegraphics[width=\textwidth, height=5cm]{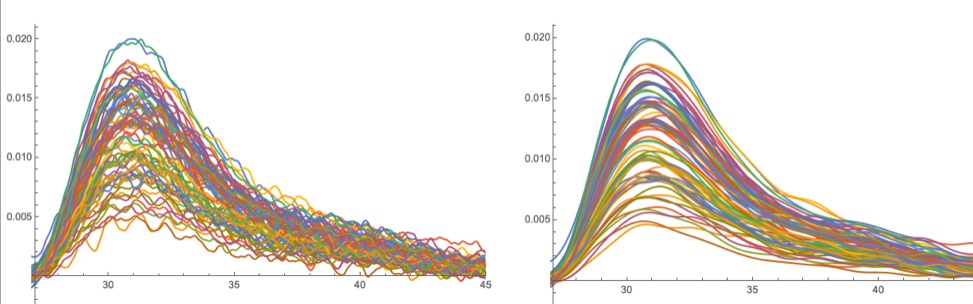}}
\caption{Typical waveforms at a mean signal level of $\sim 8$ photoelectrons (volts, nanoseconds). In the right hand plot the signals are preprocessed with a 500MHz digital low pass filter
which effectively eliminates the $\sim 800$ MHz ambient RF noise from the environment.}
\label{fig:Bandpass}       
\end{figure}

	In order to mitigate the timing degradation due to dark count noise it is planned for the next iteration of the TOFHIR ASIC\cite{tahereh} to implement an active high frequency baseline restoration which produces
a corrected signal:

 \begin{equation}
hf(t)=f(t)-f(t+\delta t)
\end{equation}
where $\delta$t of $\sim$0.5 nanoseconds(ns) is likely the optimal choice.

\begin{figure}
\centering
\centerline{\includegraphics[width=0.6\textwidth, height=5cm]{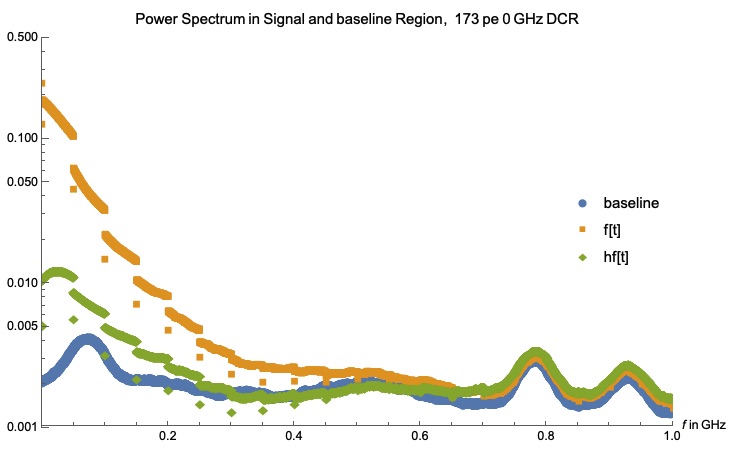}}
\caption{An example of the power spectra for the baseline/random triggers showing the prominent RF pickup around 800 MHz, the same spectrum shown for laser triggers in the signal region and finally for the signal region after
performing the subtraction in eqn.1. Note that the 500 MHz low pass filter is not applied for this figure.}
\label{fig:Bandpass}       
\end{figure}

	In contrast to the digital low pass filter discussed above, this correction essentially introduces a hi-pass filter. The combined effect of the two filters is illustrated in Figure 3 where we display
the baseline noise absent a laser signal, the original signal and the signal after applying eqn. 4.1.

	Since one aim of this note is to evaluate the effectiveness of the TOFHIR signal processing  with bench data, we report throughout time resolutions obtained using f(t)- ie the raw scope signals cleaned up with the 500 MHz low pass filter- and
also that obtained from hf(t) - as defined in eqn 4.1. 

	We use throughout the value of $\delta$t= 0.5 ns. In a section below we return to the optimal choice of $\delta$t.
	
\subsection{Signal time-of-arrival extraction}

	When we report time resolution in the following we are referring to the jitter in time-of-arrival (TOA) relative to the PicoQuant trigger time. The TOA is extracted from the waveforms in a given data set using a 
time over a threshold which is determined to be a fixed fraction of the signal peak amplitude (Constant Fraction timing). This procedure eliminates the usual ``walk correction". The interpolation between sample points is obtained
from a local polynomial fit.

	In the case of hf[t] timing, the waveforms are scaled according to the original f[t]- ie relative to the peak value of f[t] for that event. 
	
	For the results reported here we scan for the optimum value of Constant Fraction threshold and always report the best result. The optimum threshold varies rapidly with the dark count rate for the unsubtracted waveforms, f[t] ,
and is typically around $10\%$ or less of the peak value. But as the dark count rate increases it rapidly approaches $50\%$ or so. As expected for the subtracted waveforms, hf[t] , the location of the optimal threshold is less sensitive to the 
dark count rate, since the procedure of eqn 4.1 is suppressing the effect of noise.

 \section{Calibration of the number of photoelectrons-$N_{pe}$ }
 \label{sec:Nphoto}
 
	Referring to Figure 2, it is straightforward to resolve bands corresponding to the number of photoelectrons in a particular event when N$_{pe}$ is small. The cleanness of this interpretation is illustrated in Figure 4, where we have sorted events according to the pulse area rather than peak amplitude. For higher light levels there are obvious checks on the extrapolation from the few photoelectron calibration, the above mentioned calibration from the known SiPM gain and trigger repetition rate to the SiPM bias current. Because we operated the SiPM at a low 2 V over voltage the Excess Noise Factor is small and the ratio- rms divided by the Mean- of the Poisson distribution in pulse amplitude also confirms the calibration. We estimate that this procedure yields an uncertainty in N$_{pe}$ of $5\%$
or better. 

\begin{figure}
\centering
\centerline{\includegraphics[width=0.6\textwidth, height=5cm]{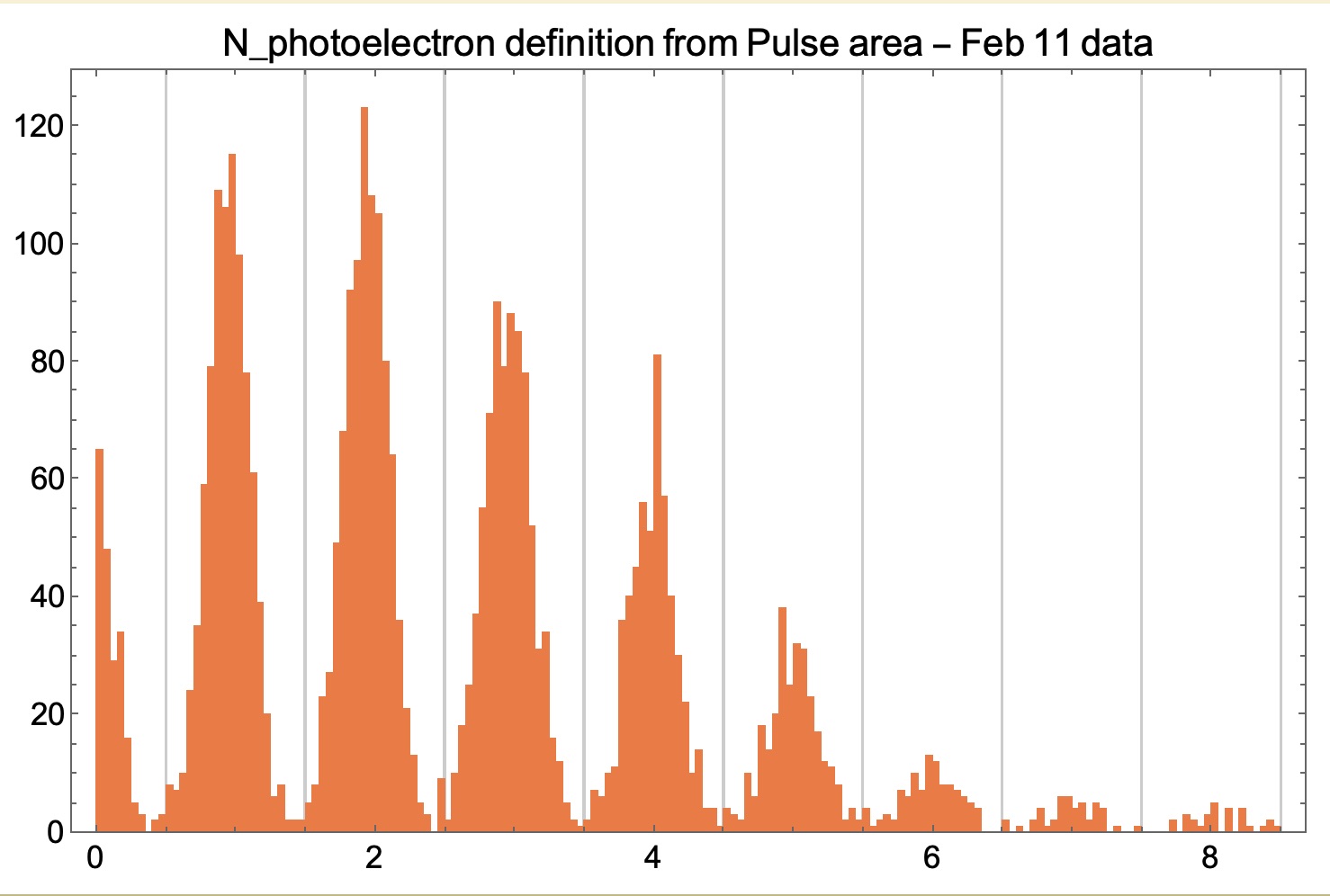}}
\caption{Original calibration set for signal amplitude vs. Number of photoelectrons.}
\label{fig:Bandpass}       
\end{figure}

\section{Stochastic term}
\label{sec:Stochastic}

	Having calibrated the signal amplitude per photoelectron we then vary the light pulse amplitude using the PicoQuant controls and analyze the waveform data for time resolution according to the above procedure. We identify this component of
the time resolution (absent dark count noise) as the ``stochastic term", shown in Figure 5. The best fit to these data yields:

 \begin{equation}
\sigma^t_{stochastic}=\frac{0.27}{\sqrt{N_{pe}}} \: \rm{ ns}
\end{equation}
 
 	This form for the stochastic term is derived from Constant Fraction timing on full waveforms- f[t]. The corresponding fit for the stochastic term differs insignificantly from this form when using the subtracted waveform- hf[t]. So we use this
	form for both cases.
	
\begin{figure}
\centering
\centerline{\includegraphics[width=0.6\textwidth, height=5cm]{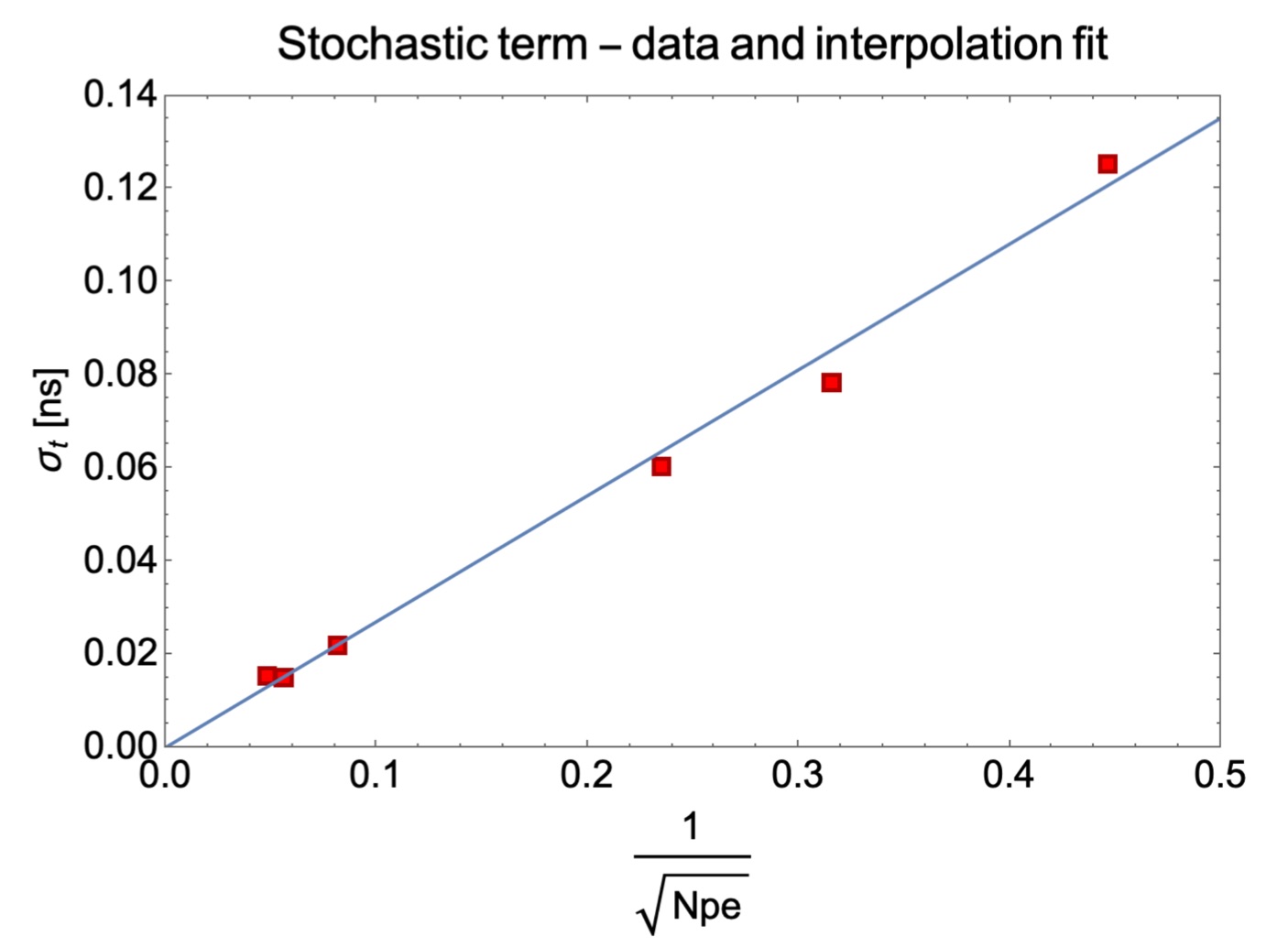}}
\caption{Stochastic term- ie the time resolution absent dark counts. The yellow and blue points were taken at the end and beginning of this series of measurements, respectively.}
\label{fig:Bandpass}       
\end{figure}

	The stochastic term is usually measured for each data set but in certain cases, ie for highly irradiated SiPMs, it had to be obtained from the above fit.
	
\section{Dark Count Rate contribution to time resolution}
\label{sec:DCR}

	To simulate the DCR in non-irradiated SiPMs we varied the current through the DC powered external 470 nm LED directly illuminating the SiPM cells uniformly. The increasing voltage drop over the resistance (2.28 k$\Omega$) in series with the SiPM vs. higher DCR was compensated so that all measurements were taken at ($2.0\pm 0.1$) V over voltage.

Figure 6 shows the bias curves after voltage compensation for different DCR between 0 and 40 GHz. The fact that the curves are parallel shows a minimum of self heating that would otherwise lower the gain due to breakdown voltage shift.
In addition we used a SiPM irradiated at a dose was $2\times 10^{12}$ n/cm$^2$  1 Mev equivalent. This corresponds to about $5\times 10^{13}$ n/cm$^2$ at -$30\deg$C which is similar to 2-3 years of operation in the real detector. In the case of a real irradiated SiPM the internal SiPM DCR generation will increase a factor of 1.85 per 10 $^{\circ}$C, yielding a slight non parallel behavior in the current vs over voltage plot vs the artificial generated DCR obtained using the LED. We calculate that the deviation of the current by $8\%$ at 2 V corresponds to a 1.3 deg C of SiPM self heating in our setup.

\begin{figure}
\centering
\centerline{\includegraphics[width=\textwidth, height=8cm]{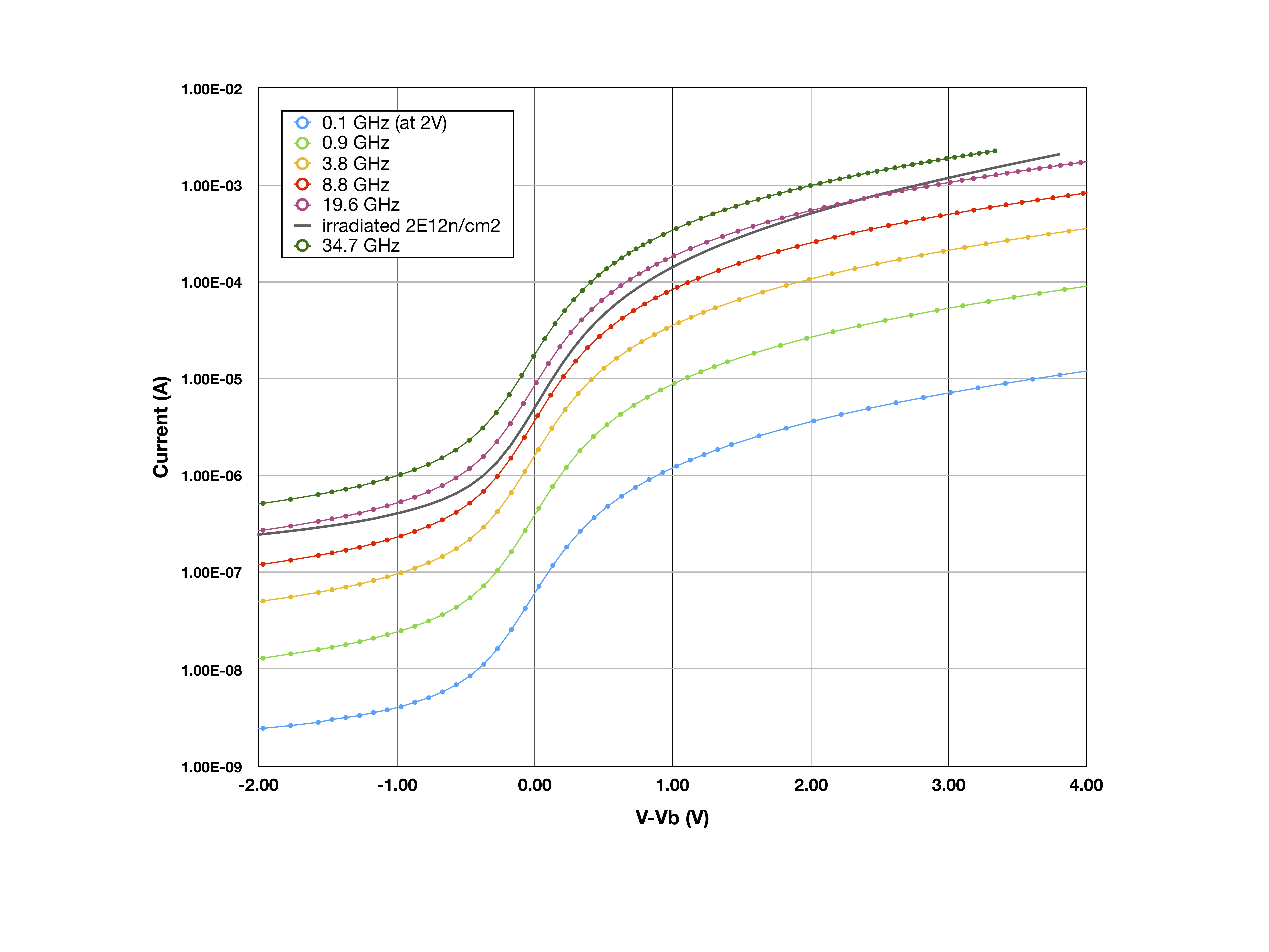}}
\caption{I-V curves corresponding to different running conditions (ie DCR ).}
\label{fig:Bandpass}       
\end{figure}

In the following we plot both the time resolution obtained from constant fraction timing on f[t] (upper points) and the corresponding values for hf[t] (lower points).
The plots themselves illustrate the $\sqrt{DCR}$ dependence of the time resolution for a given signal level. Comparing the best fits in Figures 7 and 8, we roughly confirm the
$N_{pe}^{-1}$ dependence on signal level of the DCR term. However for a more quantitive comparison we refer to the table below. We find that the hf[t] algorithm of eqn. 1
does indeed reduce the DCR term for our data by roughly a factor of 2. In both algorithms the DCR term is also roughly proportional to $N_{pe}^{-1}$ but the agreement is only
at the 15$\%$ level whereas we expect the error on $N_{pe}$ to be of order 5$\%$. It is not clear whether this discrepancy is significant.

\begin{center}
    \begin{tabular}{ | l | l | l | p{3cm} |}
    \hline
    $N_{pe}$ & coeff. of $\sqrt{DCR}$, f[t] fit&coeff. of $\sqrt{DCR}$, hf[t] fit \\ \hline
    148 &32.06 & 16.7 &    \\ \hline
    390 & 14.33 & 7.26 &  \\ \hline
   ratio & 2.24 & 2.30 &  390/148=2.64 \\
    \hline
    \end{tabular}
\end{center}

\begin{figure}
\centering
\centerline{\includegraphics[width=0.9\textwidth, height=8cm]{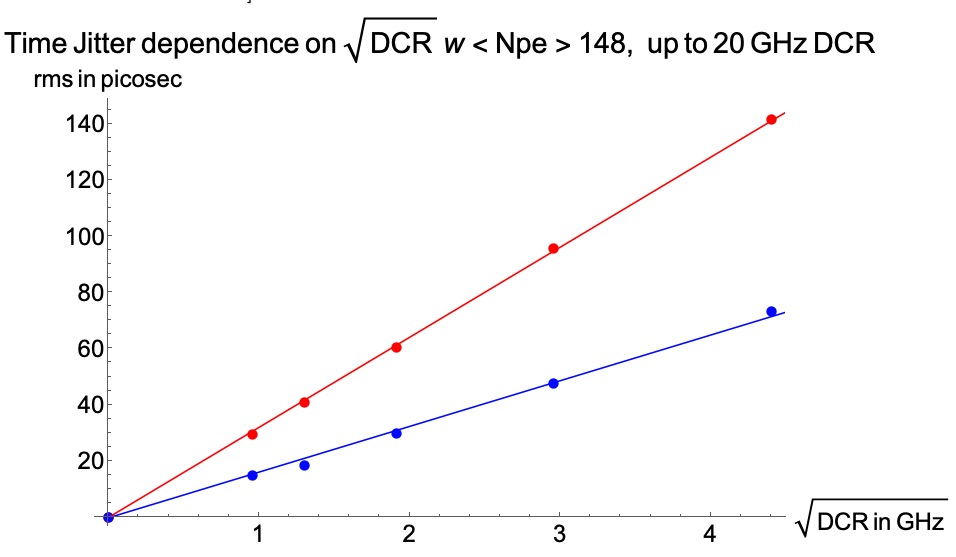}}
\caption{Data with Signal Amplitude of 148 photoelectrons mean. Points and fits correspond to the timing algorithms,
with the lower points utilizing the subtracted waveform hf[t].}
\label{fig:sum1}       
\end{figure}

\begin{figure}
\centering
\centerline{\includegraphics[width=0.9\textwidth, height=8cm]{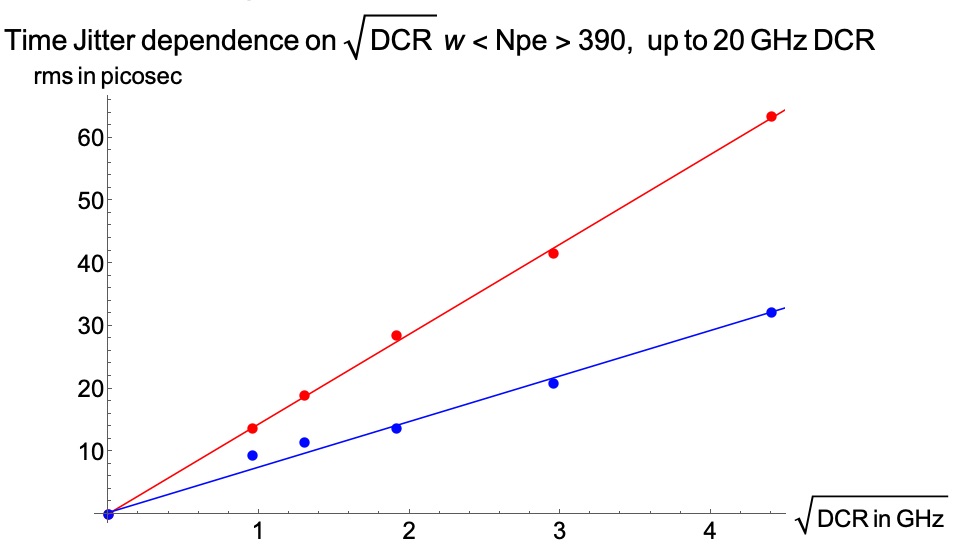}}
\caption{Data with Signal Amplitude of 390 photoelectrons mean. Points and fits correspond to the timing algorithms,
with the lower points utilizing the subtracted waveform hf[t]}
\label{fig:sum2}       
\end{figure}

	As a cross-check on this functional form we approximate the conclusions of the above table with a functional form corresponding to the best fit for row 2 (ie $N_{pe}$=390):

 \begin{equation}
\sigma^t_{DCR}=\rm{14.3 \;or\; 7.26}\times \frac{390}{N_{pe}}  \rm{\; ps- f[t] \;or\; hf[t]}
\end{equation}	
	
	We have compared this form as an extrapolation to the measured time jitter for our 34.7 GHz DCR data to the data where we used a SiPM irradiated up
to a level of $2\times10^{12} \rm{neq/cm}^2$ and and obtained good agreement in both cases. An overall summary is plotted on a linear scale in Figure 9.



\begin{figure}
\centering
\centerline{\includegraphics[width=0.7\textwidth, height=8cm]{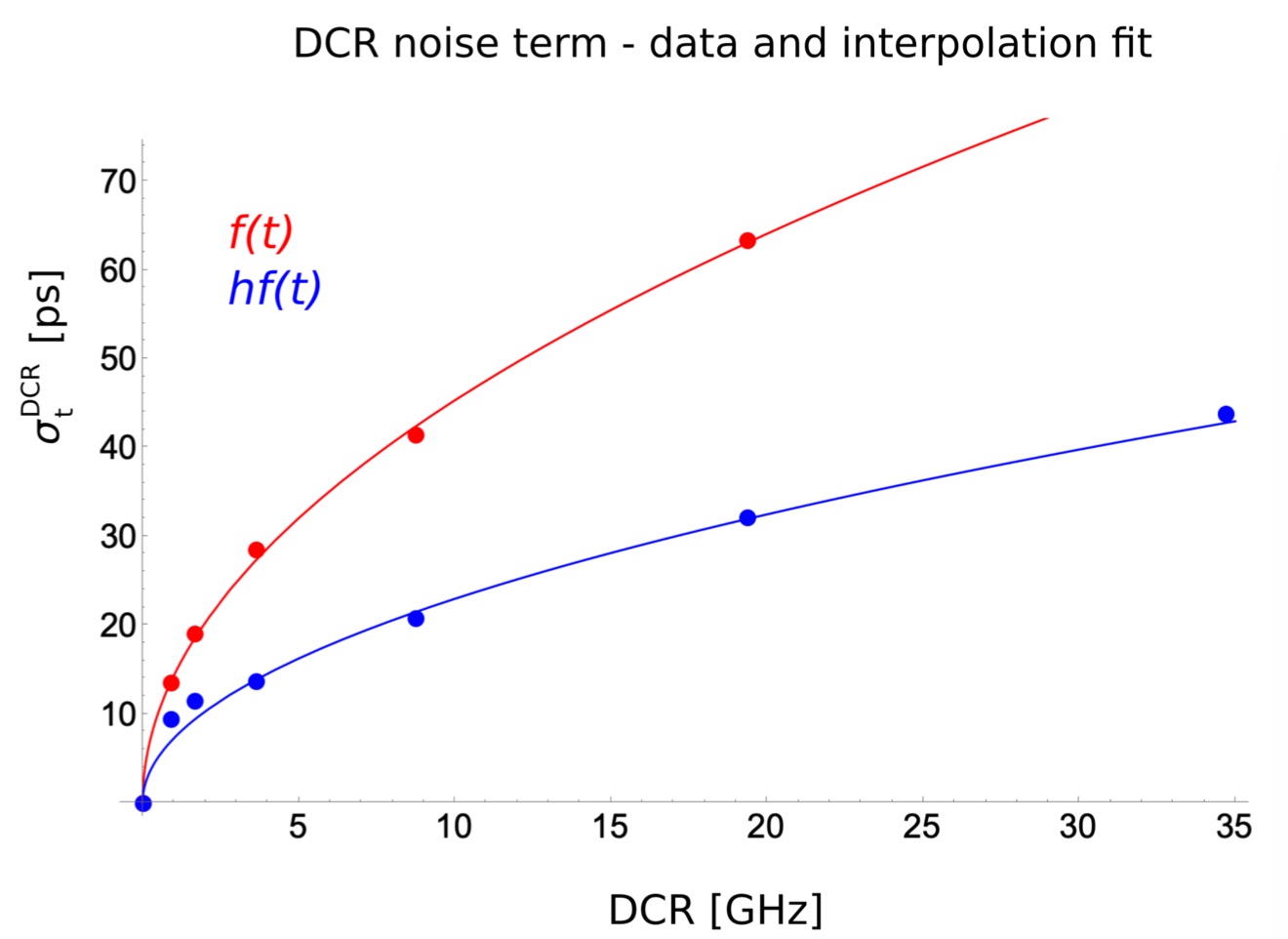}}
\caption{ Plot extended to include 40GHz (actually 34.7 GHz) using $N_{pe}$ dependence of eqn 7.1.}
\label{fig:cpirrad}       
\end{figure}

\section{ Optimization of $\delta$t- delay in the hf[t] algorithm}
\label{sec:Optdt}

	Lastly we return to the topic of possible optimization of the shift, $\delta$t, working with a representative, real case of the irradiated SiPM data. In Figure 10 we take a sample event and display the waveforms for 3 different values of dt
 (0.25, 0.5, 1.0 ns). The analyses for these data were repeated, while varying $\delta$t over this range. The resulting best fit time jitter measurements were consistent to the level of our method.
 
 	We conclude that, at least for this representative case, the time resolution is not very sensitive to the value of $\delta$t.

\begin{figure}
\centering
\centerline{\includegraphics[width=0.9\textwidth, height=4cm]{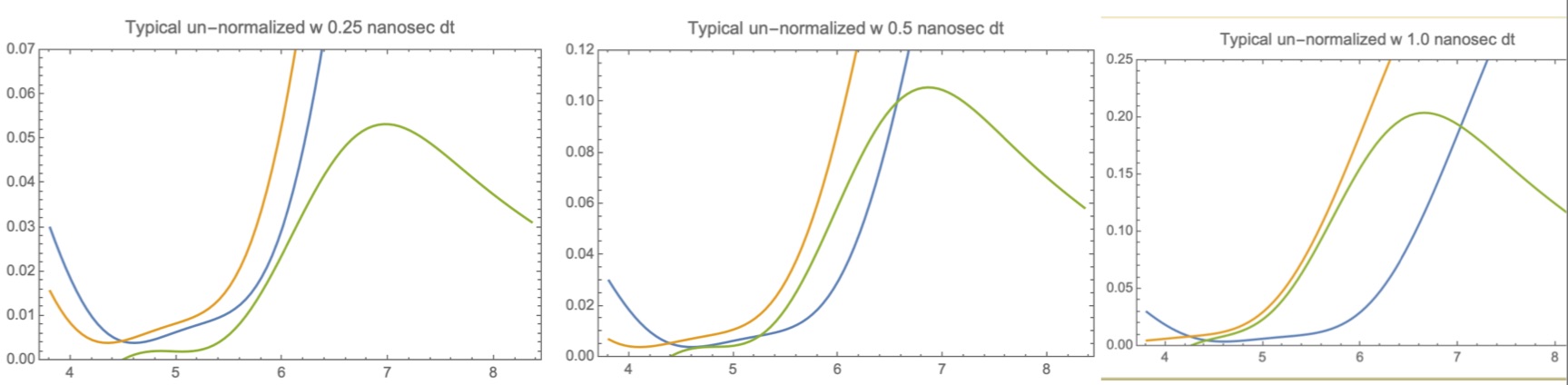}}
\caption{ A scan of hf[t] with dt=0.25 to 1.0 nanoseconds for irradiated SiPM (DCR~20 GHz) shows insignificant variation of resolution for dt in this range. The 
horizontal scale is in nanoseconds and the vertical is the amplitude (in volts) of the residual signal.}
\label{fig:cpirrad}       
\end{figure}

\subsection{Acknowledgement}
	We thank Christian Joram for making his lab and equipment available for these tests.
	This work received partial support through the US CMS program under DOE contract No. DE-AC02-07CH11359.


\begin{thebibliography}{99}

\bibitem{Uli} S. ~White, Proceedings of "ULITIMA 2018" conference, Argonne National lab, Sept.,2018. 	arXiv:1812.01425

\bibitem{Cleland} W.E. ~Cleland, E.G. ~Stern. ``Signal processing considerations for liquid ionization calorimeters in a high rate environment" Nucl. Instrum. Methods A. 1994;338:467. doi: 10.1016/0168-9002(94)91332-3.


\bibitem{intro}
S. ~White ~ "Experimental challenges of the European Strategy for Particle Physics", ~\emph{Proc. Int. Conf. on
Calorimetry for the High Energy Frontier} (CHEF 2013).

\bibitem{seb} S. ~White, ``R$\&$D for a Dedicated Fast Timing Layer in the CMS Endcap Upgrade", Proceedings of Picosecond Workshop, Clermont- Ferrand 2014, arXiv:1409.1165 [physics.ins-det].


\bibitem{tahereh} T.~ Niknejad, "First experimental results on TOFHIR readout ASIC of the CMS Barrel Timing Layer", to appear in Proceedings of TWEPP 2019.


\end{thebibliography}
\end{document}